\documentclass[twoside,slac_one]{revtex4}
\usepackage{graphicx}
\usepackage{fancyhdr}
\usepackage{amsmath} 
\usepackage{bm}
\usepackage{amsxtra}
\usepackage{amssymb}
\usepackage{amsthm}
\usepackage{latexsym}
\usepackage{lscape}

\begin{document}

\title{W and Z production in the forward region at LHCb}
\author{J. Anderson, on behalf of the LHCb collaboration.}
\affiliation{Physik-Institut der Universit\"at Z\"urich, Switzerland.}

\begin{abstract}
Measurements of electroweak boson production in pp collisions at $\sqrt{s} = 7$ TeV are presented using the decays $W \rightarrow \mu\nu$,
$Z \rightarrow \mu\mu$ and $Z \rightarrow \tau\tau$ recorded using the LHCb detector at the LHC.
For muonic $W$ and $Z$ decays, the data sample corresponds to an integrated luminosity of 37.1 $\pm$ 1.3 pb$^{-1}$. 
Here the $W$ and $Z$ bosons are reconstructed from muons with transverse momenta, $p_{T}$, above 20 GeV and pseudorapidity,
$\eta$, between 2 and 4.5, and, in the case of the $Z$, a dimuon invariant mass $M_{Z}$ between 60 GeV and 120 GeV. For $Z$
decays to tau lepton final states, 247 pb$^{-1}$ of data has been used. Here one tau is identified through its decay to a muon 
and neutrinos; the other through its decay to an electron or muon and neutrinos. The quoted $Z\rightarrow\tau\tau$ cross-section corresponds to the same fiducial volume as the $Z\rightarrow\mu\mu$ measurement, i.e. both taus must have $p_{T}$ greater than 20 GeV, pseudorapidities between 2 and 4.5 and combine to give an invariant mass between 60 and 120 GeV. The cross-sections are measured to be:
$\sigma(W^{+}\rightarrow\mu^{+}\nu) = $ 808 $\pm$ 7 $\pm$ 28 $\pm$ 28 pb; $\sigma(W^{+}\rightarrow\mu^{+}\nu) = $ 634 $\pm$ 7 $\pm$ 21 $\pm$ 22 pb;
$\sigma(Z\rightarrow\mu\mu) = $ 74.9 $\pm$ 1.6 $\pm$ 3.8 $\pm$ 2.6 pb; $\sigma(Z\rightarrow\tau\tau) = $ 82 $\pm$ 8 $\pm$ 7 $\pm$ 4 pb. 
Here the first error is statistical, the second is systematic and the third is due to 
the luminosity determination. For muonic final states, differential measurements, cross-section ratios and the $W$ charge asymmetry are also
measured in the same kinematic region. The ratio of the $Z\rightarrow\tau\tau$ and $Z\rightarrow\mu\mu$ cross-sections has been measured to be
1.09 $\pm$ 0.17, consistent with lepton universality. Theoretical predictions, calculated at next-to-next-to-leading order (NNLO) in QCD using recent
parton distribution functions, are found to be in agreement with the measurements.

\end{abstract}

\maketitle

\thispagestyle{fancy}

\section{Introduction}
LHCb \cite{LHCb}, one of the four large experiments at the Large Hadron Collider (LHC), has been primarily designed and built to make measurements of CP-violating and rare decays in the b-quark sector. Due to the $b\bar{b}$ production topology at the LHC, whereby both B hadrons are mostly produced in the same forward or backward cone, LHCb has been constructed as a forward single-arm spectrometer with an approximate coverage in terms of pseudorapidity of $1.9<\eta<4.9$. While a portion of this pseudorapidity range ($1.9<\eta<2.5$) is also covered by the general purpose detectors ATLAS and CMS, the very forward region ($\eta>2.5$) is unique to LHCb. In addition to its main B physics programme, LHCb is capable of making precision electroweak measurements at high rapidities testing theoretical predictions and enabling the exploration of a large, previously unmeasured, kinematic region. 

LHCb has been taking pp collision data at $\sqrt{s} = $ 7 TeV since March 2010. The $Z \rightarrow \mu\mu$ and $W \rightarrow \mu\nu$ cross-sections presented here are based on the full dataset collected during 2010, corresponding to an integrated luminosity of 37.1 $\pm$ 1.3 pb$^{-1}$ \cite{WZconf}. The cross-sections are measured in a fiducial region corresponding to the kinematic coverage of the LHCb detector where the final state muons have a transverse momentum, $p_{T}$, exceeding 20 GeV and have pseudorapidities in the range $2 < \eta < 4.5$. In addition, the muons from Z boson decays must combine to a dimuon invariant mass $M_{\mu\mu}$ in the range $60 \leq M_{\mu\mu} \leq 120$ GeV. The cross-sections are also measured differentially in boson rapidity and transverse momentum and lepton pseudorapidity and results are presented for the cross-section ratios $R_{WZ} = \sigma_{W}/\sigma_{Z}$ and $R_{W} = \sigma_{W^{+}}/\sigma_{W^{-}}$ and the W charge asymmetry $A_{W} = (\sigma_{W^{+}} - \sigma_{W^{-}})/(\sigma_{W^{+}} + \sigma_{W^{-}})$. The $Z\rightarrow\tau\tau$ cross-section has been measured using 247 pb$^{-1}$ of data \cite{tautauconf}, corresponding to the full 2010 dataset and part of the 2011 dataset. The quoted $Z\rightarrow\tau\tau$ cross-section corresponds to the same fiducial volume as the $Z\rightarrow\mu\mu$ measurement, i.e. both taus must have $p_{T}$ greater than 20 GeV, pseudorapidities between 2 and 4.5 and combine to give an invariant mass between 60 and 120 GeV. The data are compared to NNLO theoretical predictions obtained using recent parameterisations of the proton parton distribution functions (PDFs).

\section{Electroweak boson production at LHCb}
\begin{figure}
\centerline{\includegraphics[width=0.4\textwidth]{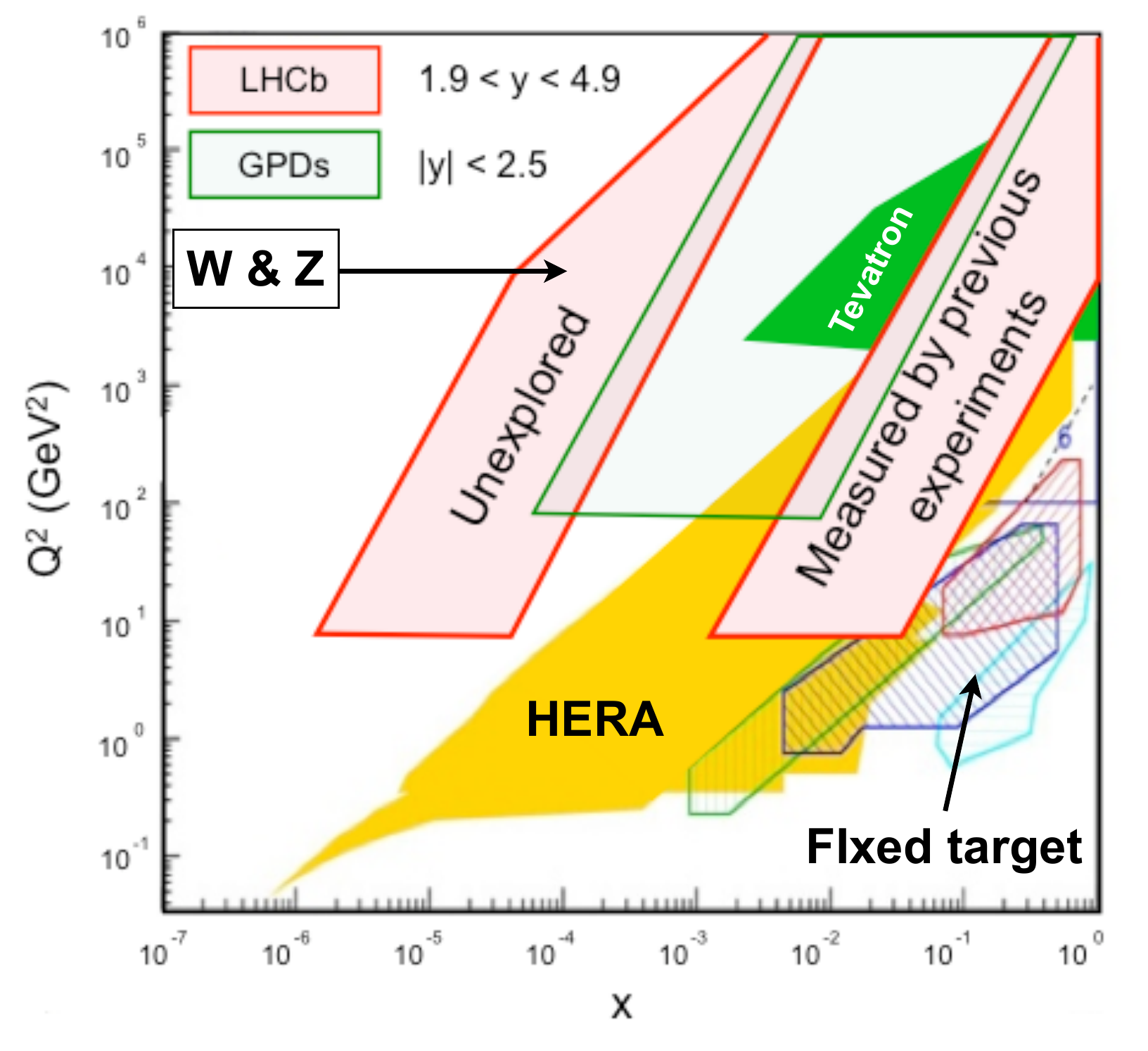}}
\caption{The kinematic range in $x-Q^{2}$ space probed at LHCb and the general purpose LHC detectors (GPDs). The regions covered by measurements at previous experiments are shown  \label{fig1}}
\end{figure}

Figure \ref{fig1} shows the kinematic region probed by events at LHCb in terms of the longitudinal fraction of the incoming proton's momentum carried by the interacting parton, x, and the square of the four-momentum exchanged in the hard scatter, Q$^{2}$. For particle production processes at LHCb, the momenta of the two interacting partons will be highly asymmetric, meaning that events at LHCb will simultaneously probe a region at high-x and a currently unexplored region at very low-x. $W$ and $Z$ production, having simple and distinctive final states, provide an ideal laboratory for investigating this unexplored region. The main theoretical uncertainties on cross-section predictions for electroweak boson production at the LHC stem from the level of knowledge of the input proton PDFs. From the point of view of LHCb, PDFs have been determined from fixed target data and to a lesser extent HERA data for the larger x values and confirmed at higher Q$^{2}$ by $W$ and $Z$ production at the Tevatron. For the smaller x values, the PDFs have been measured by HERA alone but at much lower Q$^{2}$ from where they must be evolved to higher energies using the DGLAP equations. 

\begin{figure}
\centerline{\includegraphics[width=0.8\textwidth]{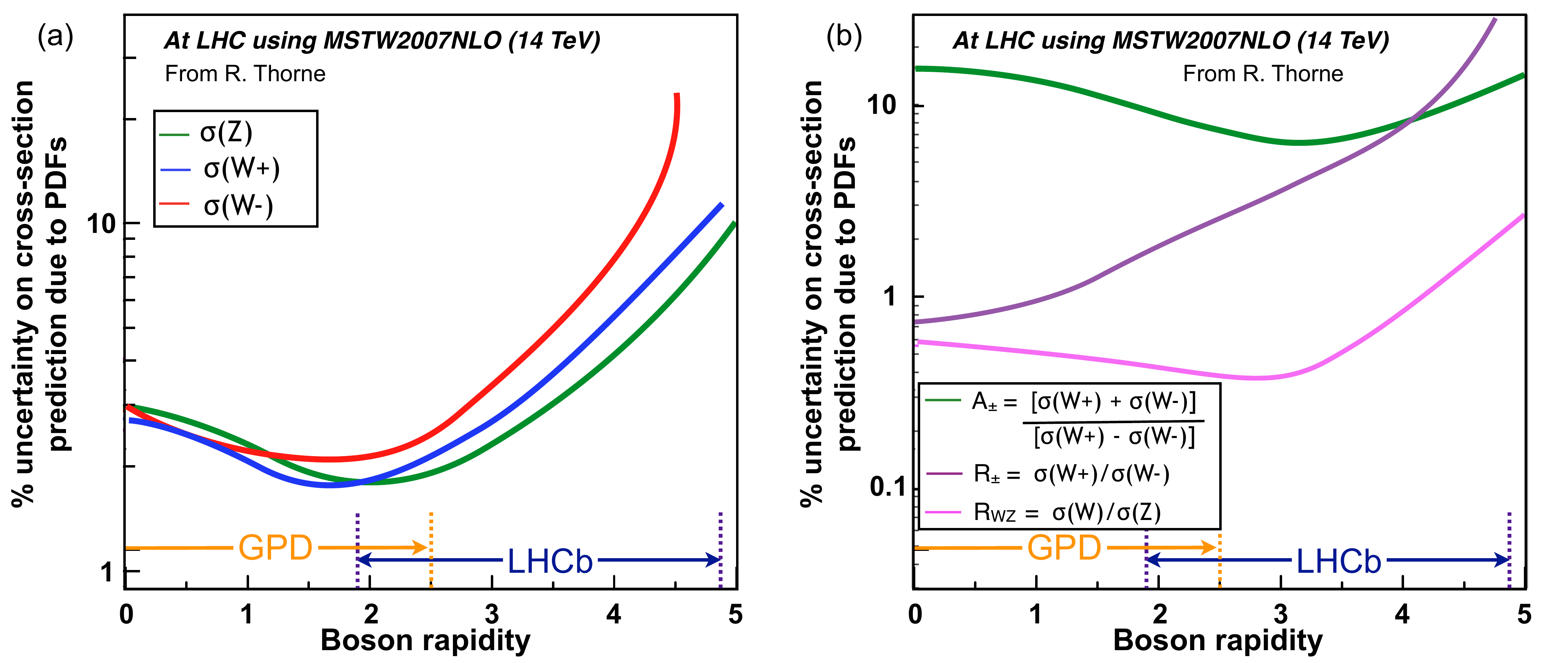}}
\caption{(a) Percentage uncertainty on cross-section predictions for W, Z and low mass Drell-Yan pairs at the LHC due to the PDFs as a function of
  rapidity. The regions fully instrumented by LHCb and the GPDs are shown (from \cite{MSTW07}). (b) Percentage uncertainty on predictions for the ratio and asymmetry of W and Z production at the LHC due to the PDFs. (from \cite{MSTW07}). \label{fig2}}
\end{figure}

Measurements of the differential cross-sections for electroweak bosons at LHCb and the ratios of these cross-sections can provide useful information on both the PDFs and QCD. While electroweak theory can currently describe the fundamental partonic processes of electroweak boson production at the LHC at NNLO with an accuracy at the percent level, a large uncertainty on the theoretical predictions arises from the present knowledge of the proton PDFs. Figure \ref{fig2}(a) shows the percentage uncertainty on cross-section predictions for $W$ and $Z$ production at the LHC due to the uncertainty on the PDFs. These uncertainties are largest for high boson rapidities; consequently measurements at LHCb, which is fully instrumented in the forward region, can provide input to constrain the PDFs, both in the unique forward region with pseudorapidities $\eta > 2.5$ and in the region which is common to ATLAS and CMS ($2 < \eta < 2.5$). In addition, as shown in Figure \ref{fig2}(b) a measurement of both the ratio and asymmetry of W$^{+}$ to W$^{-}$ production can provide further constraints on the PDFs while a measurement of the ratio of W to Z production, being almost unaffected by PDF uncertainties, will provide a precision test of the Standard Model. The ratios and charge asymmetry can be measured with a higher precision because experimental and theoretical uncertainties partially cancel. The W charge asymmetry, being sensitive to the valence quark distribution in the proton, is of particular interest since it provides complementary information to the results from deep-inelastic scattering cross-sections at HERA as the HERA data does not strongly constrain the ratio of $u$ over $d$ quarks at low x.   

\section{Selection of $Z \rightarrow \mu\mu$ Candidates}
\begin{figure}
\centerline{\includegraphics[width=0.5\textwidth]{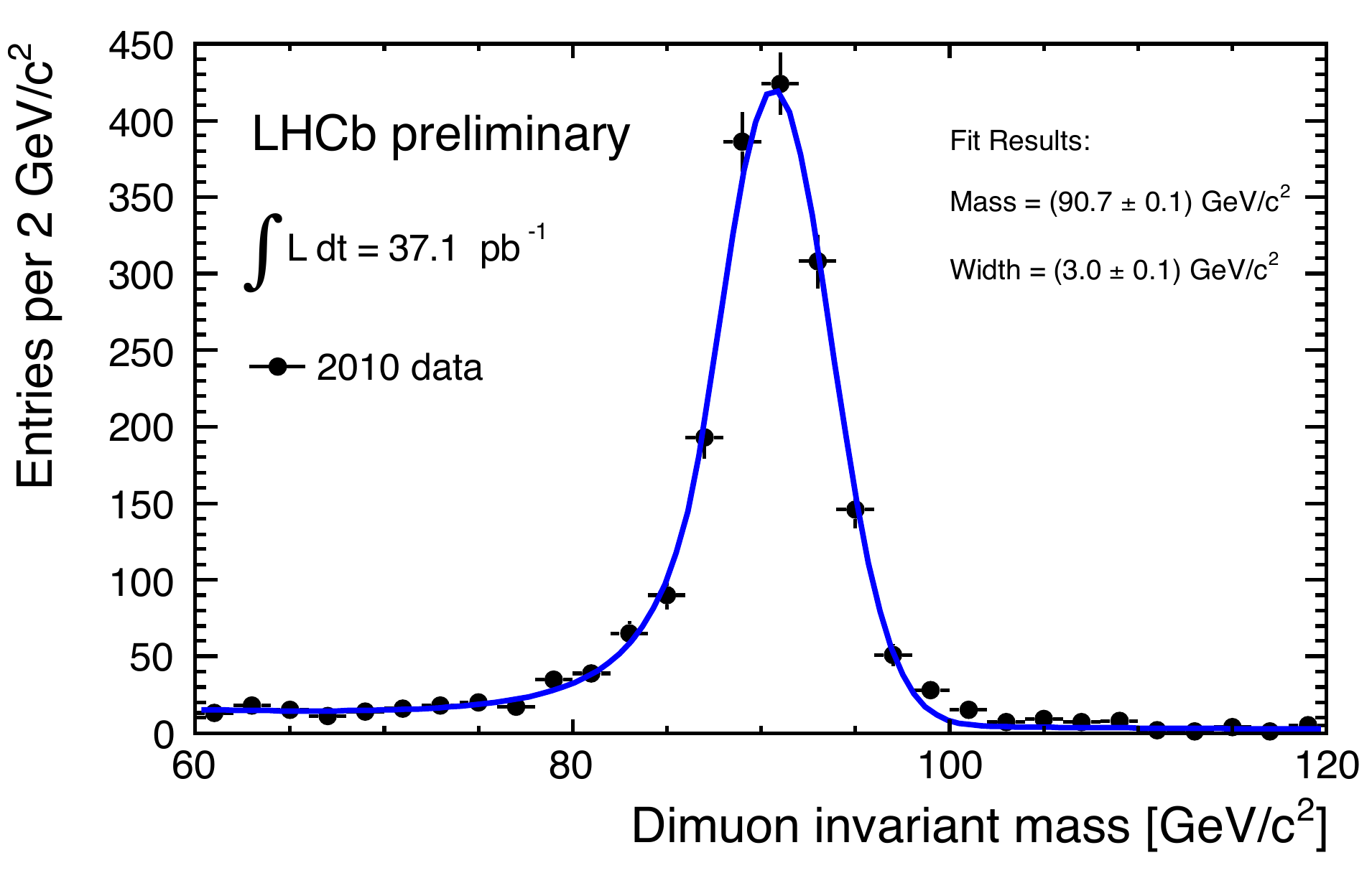}}
\caption{Dimuon invariant mass distribution of the selected Z boson candidates. The data are fitted with a Crystal Ball function for the signal and an exponential for the Drell-Yan dimuon production and the background contribution. \label{fig3}}
\end{figure}

$Z \rightarrow \mu\mu$ candidates are initially selected at LHCb via a single muon trigger that requires the presence of at least one muon that has a transverse momentum greater than 10 GeV. Additionally, global event cuts (GECs) on the tracker hit multiplicity are applied in the trigger to remove very high multiplicity events. Offline the events are identified by requiring a pair of oppositely charged muons which combine to a dimuon invariant mass in the range $60 \leq M_{\mu\mu} \leq 120$ GeV. Both muons are required to have a good track quality, transverse momenta larger than 20 GeV and pseudorapidities between 2 and 4.5 and an associated hit in each of the 5 muon stations. Using a dataset of $37.1 \pm 1.3$ pb$^{-1}$, 1966 Z candidates are identified. The reconstructed dimuon invariant mass distribution is shown in figure \ref{fig3}. 

The background contamination in this sample is very low. The following backgrounds have been evaluated: 

\begin{itemize}
\item $Z \rightarrow \tau\tau$ where both taus decay leptonically to muons and neutrinos. The background from this source has been estimated from simulation with the Z cross-section fixed to the cross-section measured in this analysis, the contribution from this source is $0.61 \pm 0.04$ events. 

\item Backgrounds coming from semi-leptonic decays of hadrons containing b and c quarks. This background has been evaluated from data by investigating a region that is background enhanced by requiring that both muons have impact parameters that are inconsistent with the primary vertex. The Monte-Carlo prediction from PYTHIA \cite{PYTHIA} is then used to extrapolate this value to the total contribution from this source of $4.3 \pm 1.7$ events. 

\item Backgrounds from hadron mis-identification have also been estimated from data. Since this background source should cause equal numbers of same-sign and opposite-sign Z candidates, the background contamination can be estimated by examining a same-sign sample. No Z candidates are found in this sample and a background contamination of $0 \pm 1$ events is estimated. The total background contamination in the Z sample is therefore $4.9 \pm 2.0$ events.
\end{itemize}  

\section{Selection of $W \rightarrow \mu\nu$ Candidates}
$W \rightarrow \mu\nu$ candidates are selected by the same single muon trigger line as the $Z\rightarrow\mu\mu$ events and must pass the GECs in the trigger. Offline the events  are identified by requiring an isolated high-$p_{T}$ muon consistent with the primary vertex satisfying tight track quality and muon identification criteria. The track must have a $p_{T}$ larger than 20 GeV and a pseudorapidity between 2 and 4.5. The track isolation criteria requires that there is less than 2 GeV of transverse energy in a cone of dimension $\Delta R = 0.5$ surrounding the muon track. Only muon candidates having an associated hit in each of the 5 muon stations and having associated calorimeter energy divided by track momentum (E/p) less than 0.05 are considered. After these requirements 15608 $W^{+}$ and 12301 $W^{-}$ candidates are observed in the $37.1 \pm 1.3$ pb$^{-1}$ dataset.

The background contamination is estimated by performing a template fit to the selected muon $p_{T}$ distribution in five bins in muon pseudorapidity (2.0-2.5; 2.5-3.0; 3.5-4.0; 4.0-4.5). The following background sources have been considered. 

\begin{itemize}
\item $\gamma^{\star}/Z \rightarrow \mu\mu$, where one of the muons goes outside the LHCb acceptance: the shape of the transverse momentum distribution is estimated from POWHEG \cite{POWHEG} with the Z cross-section fixed to that measured in this analysis. 

\item $W \rightarrow \tau\nu$, where the tau decays leptonically to a muon: the shape and relative number of $W \rightarrow \tau\nu$ to $W \rightarrow \mu\nu$ events is taken from PYTHIA  \cite{PYTHIA}. 

\item $Z \rightarrow \tau\tau$, where both taus decay to muons: the shape is taken from PYTHIA with the Z cross-section fixed to that measured in this analysis. 

\item Hadrons that punch through the calorimeters and leave hits in the muon stations: the shape is taken from a sample of muon candidates that pass the W selection criteria but have E/p greater than 0.1. 

\item Hadrons that decay in flight to muons before the calorimeters: here the $p_{T}$ distribution is modeled using a sample of high momentum tracks, assumed to be pions and kaons, that is weighted according to the probability of their decaying in flight, which is calculated from their mean lifetime and their boost. 
\end{itemize}

\begin{figure}
\centerline{\includegraphics[width=0.9\textwidth]{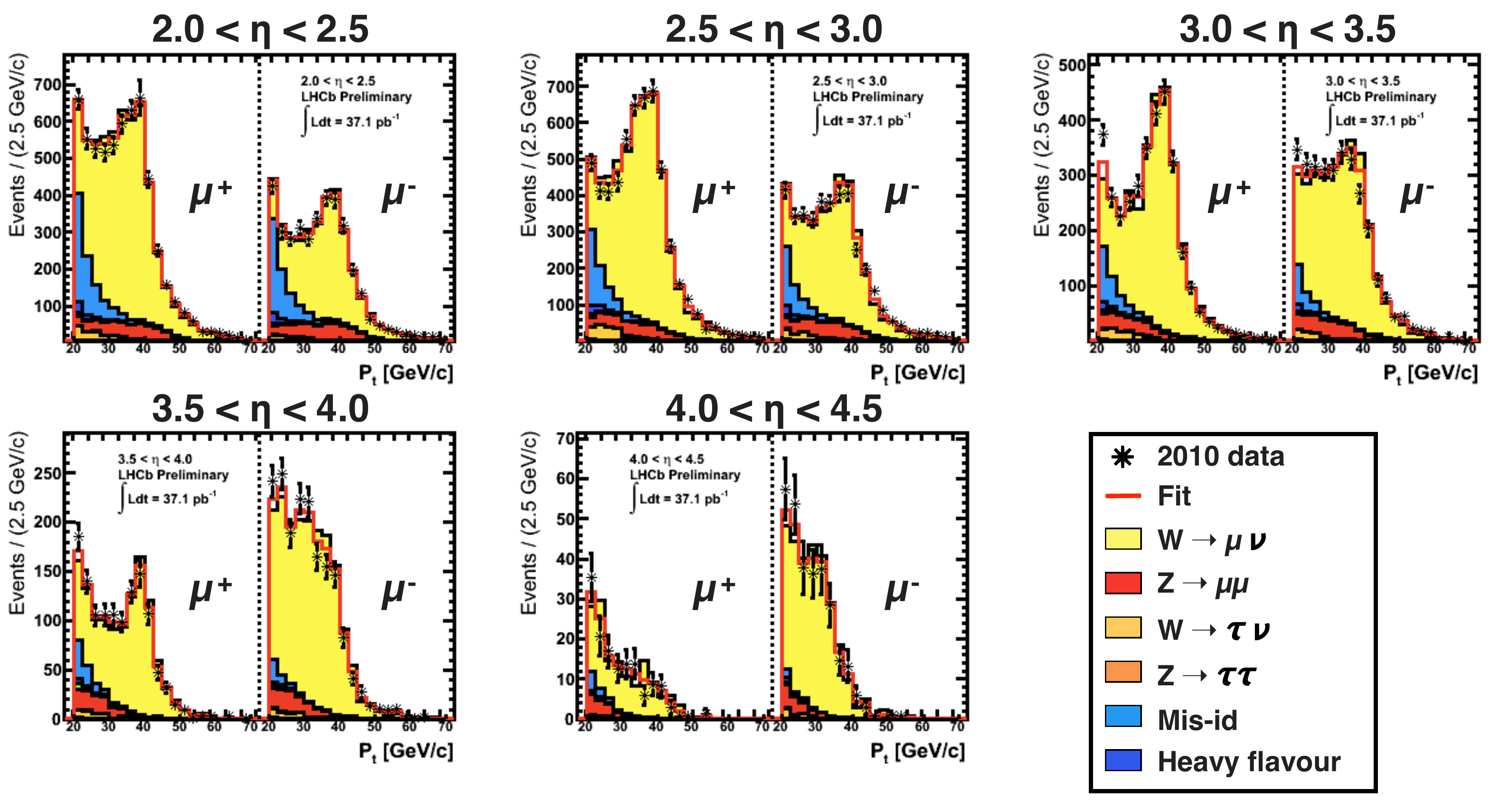}}
\caption{Muon $p_{T}$ distribution for positive (left) and negative (right) selected W candidates, in five bins of muon pseudorapidity. \label{fig4}}
\end{figure}

The shapes of the $W^{+}$ and $W^{-}$ distributions are taken from next-to-leading order simulation, using POWHEG and the CTEQ6m PDF set.  In the fit, only the overall normalisations of the signal and the backgrounds due to hadron decay in flight are allowed to vary. The fit is performed for both charges and over all pseudorapidity bins simultaneously. As a cross-check, the fit is repeated also allowing the normalisation of each pseudorapidity bin to float independently, and a similar fraction is obtained. The dominant background is due to pion and kaon decays in flight. The fitted muon transverse momenta distributions for each bin in muon pseudorapidity are shown in Figure \ref{fig4}. The overall sample purities for $W^{+}$ ($W^{-}$) are determined to be 80\% (78\%).

The efficiency of the offline $W\rightarrow\mu\nu$ selection is determined from data by using an offline selected $Z\rightarrow\mu\mu$ events where one of the muons is ignored. Depending on the pseudorapidity bin in question, an efficiency of between 45\% and 80\% is measured.

\section{Selection of $Z \rightarrow \tau\tau$ Candidates}
For the measurement of the $Z\rightarrow\tau\tau$ cross-section, two different final states have been considered: one where both tau leptons decay to muons and neutrinos and one where one tau decays to a muon and neutrinos and the other decays to an electron and neutrinos. In both channels, a muon with a $p_{T}$ greater than 20 GeV and pseudorapidity between 2 and 4.5 is required. Since both channels contain a high $p_{T}$ muon, the events are again selected by a single muon trigger requiring a muon with $p_{T}$ greater than 10 GeV. The events must also pass the GECs in the trigger.

In the $e\mu$ channel an electron having $p_{T}$ above 5 GeV, pseudorapidity between 2 and 4.5, associated energy in the electromagnetic calorimeter over the track momentum (E$_{ecal}$/p) greater than 0.1 and associated energy in the hadronic calorimeter over track momentum (E$_{hcal}$/p) less than 0.05 is required while in the $\mu\mu$ channel a second muon is required with a $p_{T}$ greater than 5 GeV and pseudorapidity between 2 and 4.5. The same selection criteria are applied to both channels to suppress QCD, top and  electroweak diboson (WW, WZ, ZZ) backgrounds. Since the two leptons produced in diboson and top production are not strongly correlated in the transverse plane, the difference between the azimuthal angles of the leptons, $\Delta\phi$, is required to be larger than 2.7 radians. The QCD backgrounds are suppressed by requiring that both lepton tracks are isolated. In the $\mu\mu$ channel there is one other significant background: oppositely charged muon pairs coming from $\gamma^{\star}/Z\rightarrow\mu\mu$ decays. Three cuts have been applied to suppress this background. Firstly, in order to remove the Z peak, the reconstructed dimuon invariant mass is required to be less than 80 GeV. Secondly, since the transverse momentum of the two muons is less balanced for the signal than the background, a cut is applied to the momentum balance between the final state muons, $p^{balance}_{T} = (p^{(1)}_{T} - p^{(2)}_{T})/(p^{(1)}_{T} + p^{(2)}_{T}) > 0.2$. Thirdly, since the tau lepton has a significant lifetime, the two muons are required to be inconsistent with the primary vertex. Finally, in order to suppress QCD backgrounds, a track isolation requirement is placed on both muons.

\begin{figure}
\centerline{\includegraphics[width=0.9\textwidth]{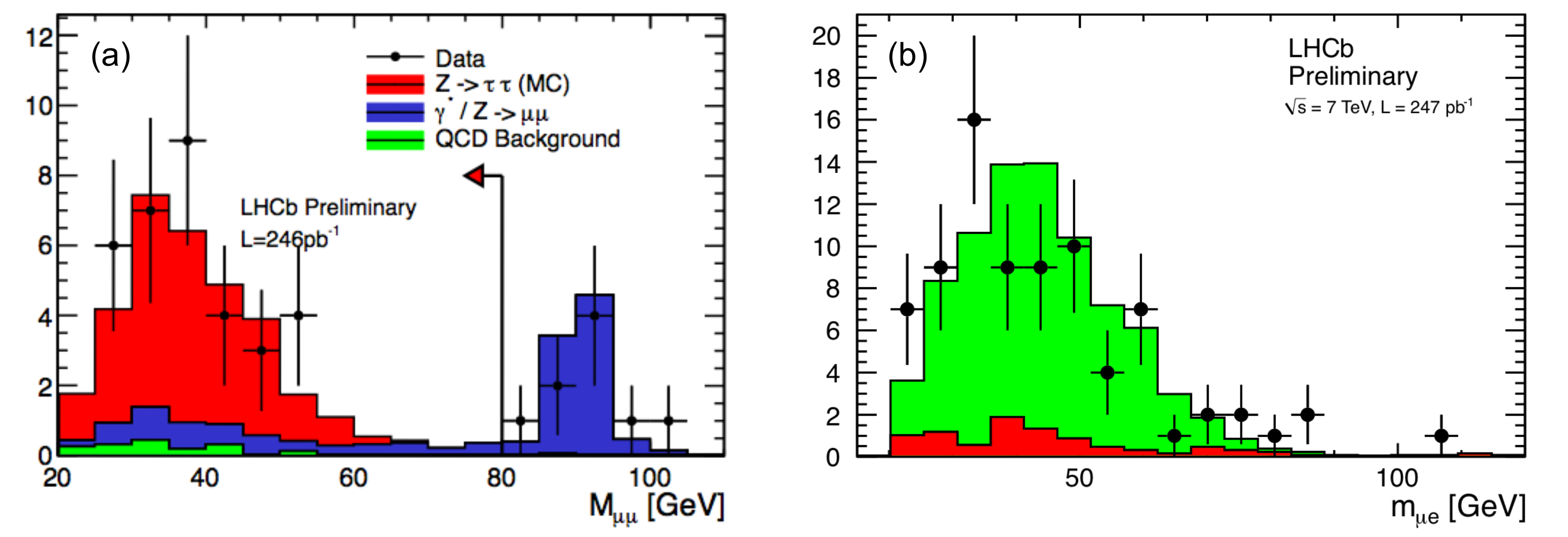}}
\caption{Invariant mass of the dileptons in the $\mu\mu$ (left) and $e\mu$ (right) channels. The points are data while the solid histograms show the expected background contamination and the expected signal from simulation. \label{fig5}}
\end{figure}

Figure \ref{fig5} shows the distribution for the invariant mass of the two leptons in each channel after the selection is applied. The estimated background contribution is also shown. The electroweak diboson backgrounds have been taken from simulation while, for the $\mu\mu$ channel, the $\gamma^{\star}/Z\rightarrow\mu\mu$ background is estimated from data by normalising the expected dimuon mass shape to the number of events with dimuon invariant masses above 80 GeV, where the sample is dominated by $\gamma^{\star}/Z\rightarrow\mu\mu$ events. For both channels, the QCD background has been obtained from data by relaxing the requirements on the lepton identification and using a sample of same-sign events normalized to non-isolated opposite-sign events. After the election cuts are applied there are 81 candidate events in the $e\mu$ channel with an expected background contamination of 12.5 $\pm$ 4.5 events while in the $\mu\mu$ channel 33 candidates are selected with a background contamination of 7.1 $\pm$ 3.6 events.

The selection efficiency in the $e\mu$ analysis has been determined from the $Z\rightarrow\tau\tau$ simulation corrected using the agreement between $Z\rightarrow\mu\mu$ data and simulation. Thus,
$\epsilon^{Z\rightarrow\tau\tau}_{sel}$(MC) as determined in the simulation is scaled by $\epsilon^{Z\rightarrow\mu\mu}_{sel}$(data)/$\epsilon^{Z\rightarrow\mu\mu}_{sel}$(MC) where the same selection requirements have been imposed on $Z\rightarrow\mu\mu$ simulation and data. A value of 46 $\pm$ 3\% is found. The systematic uncertainty is calculated by combining in quadrature the differences for each cut between efficiencies calculated using $Z\rightarrow\mu\mu$ events in data and simulation.

The $\mu\mu$ analysis has additional requirements on the muon momentum balance and the compatibility of the track with the primary vertex. However, since these cuts are designed to remove $Z\rightarrow\mu\mu$ events, it is not possible to use the remaining number of $Z\rightarrow\mu\mu$ in data and simulation to calculate a systematic. Consequently, for both these variables, anti-cuts are applied to the control sample requiring $p^{balance}_{T}$ and tracks compatible with the primary vertex. The difference between the efficiencies in $Z\rightarrow\mu\mu$ data and simulation for each selection variable are added in quadrature resulting in a selection efficiency of 17.2 $\pm$ 1.4\%.

Unlike in the case of the $W\rightarrow\mu\nu$ and $Z\rightarrow\mu\mu$ analyses where the selection criteria define the fiducial region of the measurement, the acceptance for the $Z\rightarrow\tau\tau$ measurement is not equal to one. In order to compare directly to the $Z\rightarrow\mu\mu$ measurement, we choose to quote the result in the same kinematic region, requiring both leptonic products of the Z (in this case taus) to be within pseudorapidities of 2 and 4.5, to have transverse momenta above 20 GeV, and an invariant mass between 60 and 120 GeV. Since we actually measure the decay products of the taus and do not detect them directly, we require an acceptance factor, A, in order to correct from the kinematic range in which the final states can be observed. This number has been evaluated to be 24.9 $\pm$ 1.2\% for the $e\mu$ analysis and 38.6 $\pm$ 0.9\% for the $\mu\mu$ analysis. The larger uncertainty for the $e\mu$ analysis comes from the precision with which we estimate the energy loss for the electron as it interacts with the material (which presents about 30\% of a radiation length) before the last tracking station. The uncertainty for the $\mu\mu$ analysis comes from the difference in the acceptance as calculated using PYTHIA \cite{PYTHIA} and HERWIG \cite{HERWIG}.

\section{Reconstruction Efficiencies}
For the $Z\rightarrow\mu\mu$ and $W\rightarrow\mu\nu$ analyses the efficiencies for triggering, muon identification and track finding have all been estimated from data. The trigger efficiency has two components, one due to the efficiency of the single muon trigger and the other one due to global event cuts (GEC) applied in the trigger to remove very high multiplicity events. The single muon trigger efficiency has been determined using an offline selected Z sample. By requiring that one of the muons in the event caused the single muon line to fire the other muon can be used to determine the trigger efficiency. The GEC efficiency has been determined as a function of the number of primary vertices in the bunch crossing by adding randomly triggered events to Z candidate events containing only one primary vertex. The average GEC efficiency for W and Z events amounts to 94.7 $\pm$ 0.2\%. The total trigger efficiency for Z (W) events is determined to be 90.2 $\pm$ 0.5 \% (78.0 $\pm$ 0.5\%). The track finding and muon identification efficiencies are measured using a tag and probe method and an offline selected Z sample where the tag muon is required to have fired the single muon trigger line. Since different track quality criteria are applied to the muon candidates, the single track finding efficiency is different for W and Z events and is measured to be 79.0 $\pm$ 3.1\% and 81.5 $\pm$ 3.1\% respectively. The single muon identification efficiency has been determined to be 97.9 $\pm$ 0.7\%.

The trigger, muon identification and muon tracking efficiencies determined from data can be reused for the $Z\rightarrow\tau\tau$ analysis. Since the energy losses for muons and electrons, the electron tracking efficiency has been taken from simulation scaled by the ratio of the muon tracking efficiency in data to the muon tracking efficiency in the simulation, a value of 80 $\pm$ 3\% is obtained. The electron identification efficiency has been determined from data using a $Z\rightarrow ee$ sample and a tag-and-probe technique. 

\section{Results}
Within the fiducial volumes defined in the previous sections, the W and Z cross-sections are measured to be: 
\vspace{1.5mm}

$\sigma(W^{+} \rightarrow \mu^{+}\nu) = 808 \pm 7 \pm 28 \pm 28$ pb
\vspace{1.5mm}

$\sigma(W^{-} \rightarrow \mu^{-}\nu) = 634 \pm 7 \pm 21 \pm 22$ pb
\vspace{1.5mm}

$\sigma(Z \rightarrow \mu\mu) = 74.9 \pm 1.6 \pm 3.8 \pm 2.6$ pb
\vspace{1.5mm}

$\sigma(Z \rightarrow \tau\tau) = 82 \pm 8 \pm 7 \pm 4$ pb
\vspace{1.5mm}

\noindent where the first error is statistical, the second systematic and the third comes from the luminosity determination. The integrated luminosity was determined using the Van der Meer scan method and has a 3.5\% uncertainty. The measurements have been corrected for the effects of final state radiation using HORACE \cite{HORACE}. The results are compared to theoretical predictions calculated at NNLO with the DYNNLO \cite{DYNNLO} program using the NNLO PDF sets of MSTW08 \cite{MSTW08}, ABKM09 \cite{ABKM09}, JR09 \cite{JR09} and NNPDF \cite{NNPDF} and at NLO using MCFM \cite{MCFM} and the NLO PDF sets from CTEQ \cite{CTEQ}, and NNPDF \cite{NNPDF}. Within the uncertainties the $Z$ and $W^{-}$ measurements agree well with the all the NNLO predictions while the measured $W^{+}$ cross-section is below some of the theory predictions but compatible with others. The ratio of the $Z\rightarrow\mu\mu$ and $Z\rightarrow\tau\tau$ cross-sections has been measured to be 1.09 $\pm$ 0.17, consistent with lepton universality.

\begin{figure}
\centerline{\includegraphics[width=0.9\textwidth]{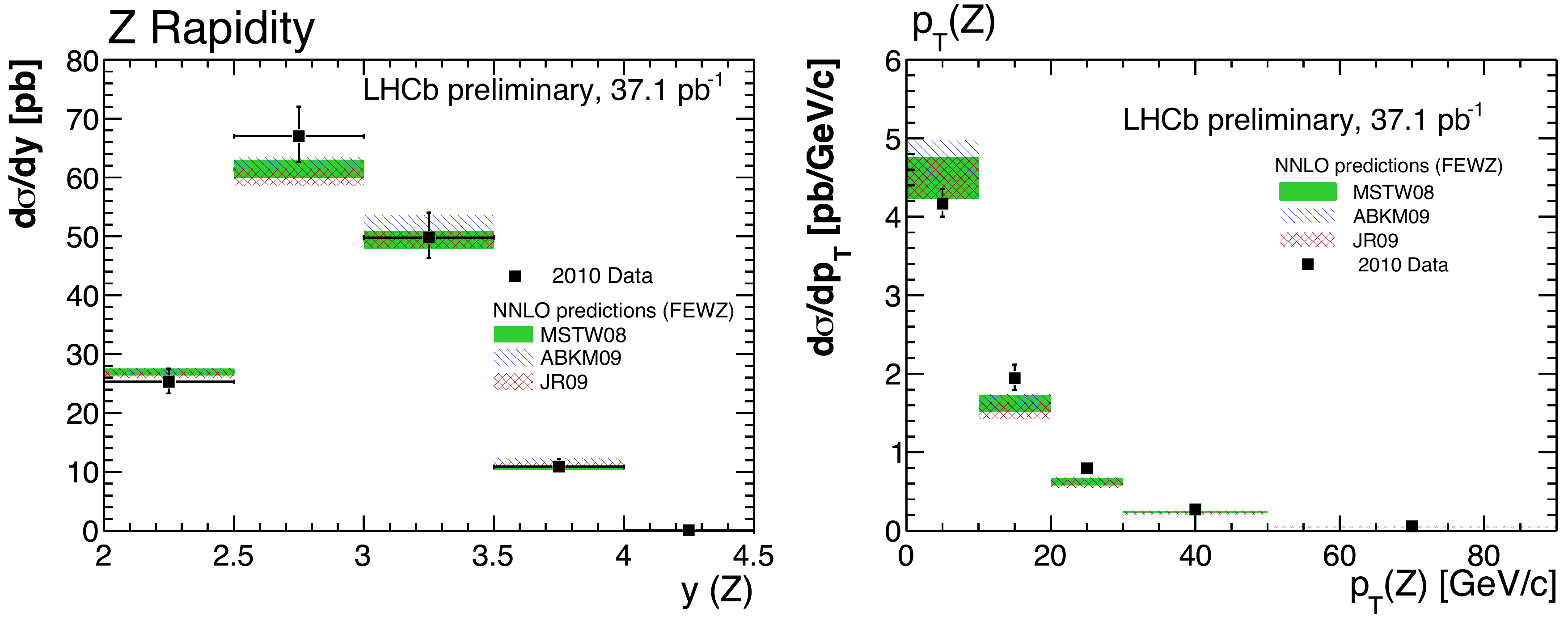}}
\caption{Differential cross-section for $Z \rightarrow \mu\mu$ as a function of Z boson rapidity, y(Z), and transverse momentum, $p_{T}$(Z). NNLO predictions using three different PDF sets are also shown. The bands show the PDF uncertainties evaluated at the 68\% confidence level. \label{fig6}}
\end{figure}

\begin{figure}
\centerline{\includegraphics[width=0.9\textwidth]{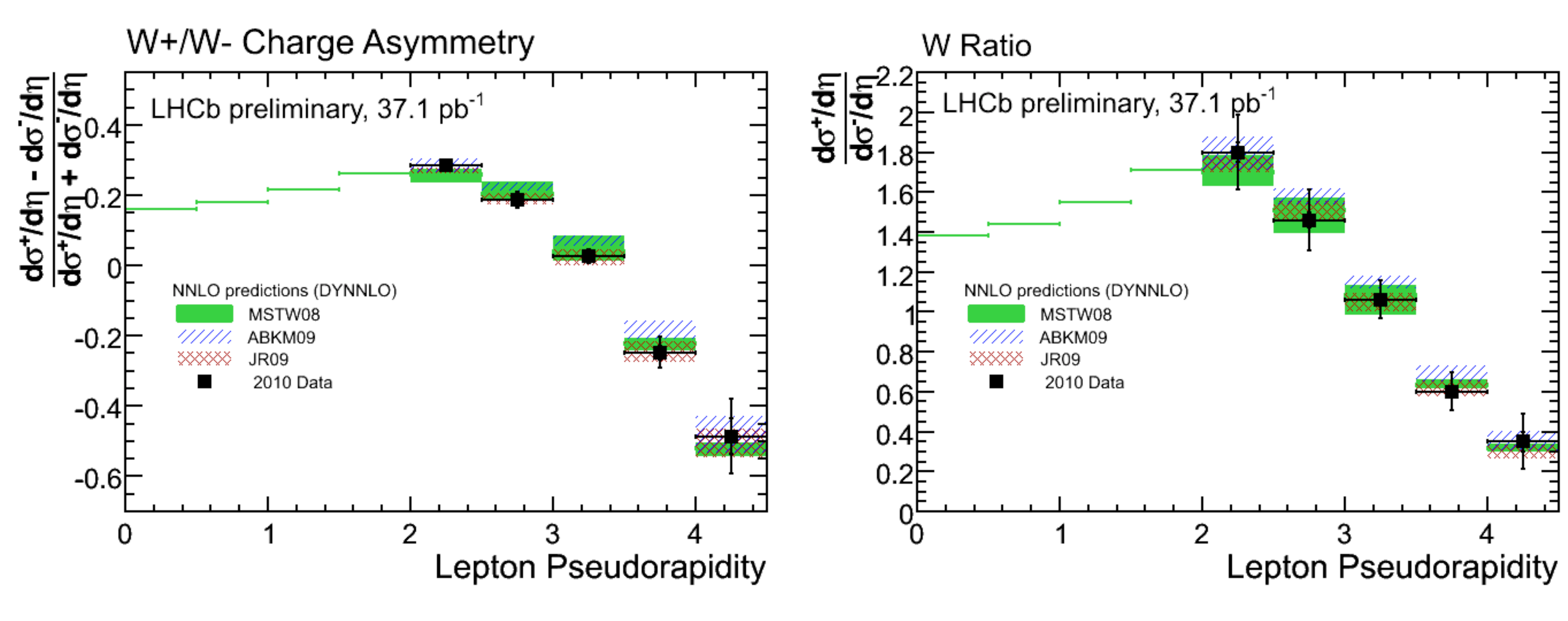}}
\caption{W charge asymmetry $A_{W} = (\sigma_{W^{+}} - \sigma_{W^{-}})/(\sigma_{W^{+}} + \sigma_{W^{-}})$ (left) and the ratio $R_{W} = \sigma_{W^{+}}/\sigma_{W^{-}}$ (right) in bins of muon pseudorapidity compared to the NNLO predictions. The shaded and hatched bands represent the uncertainty arising from the PDF sets. The line represents the central value of the prediction for pseudorapidities below 2.  \label{fig7}}
\end{figure}

\begin{figure}
\centerline{\includegraphics[width=0.5\textwidth]{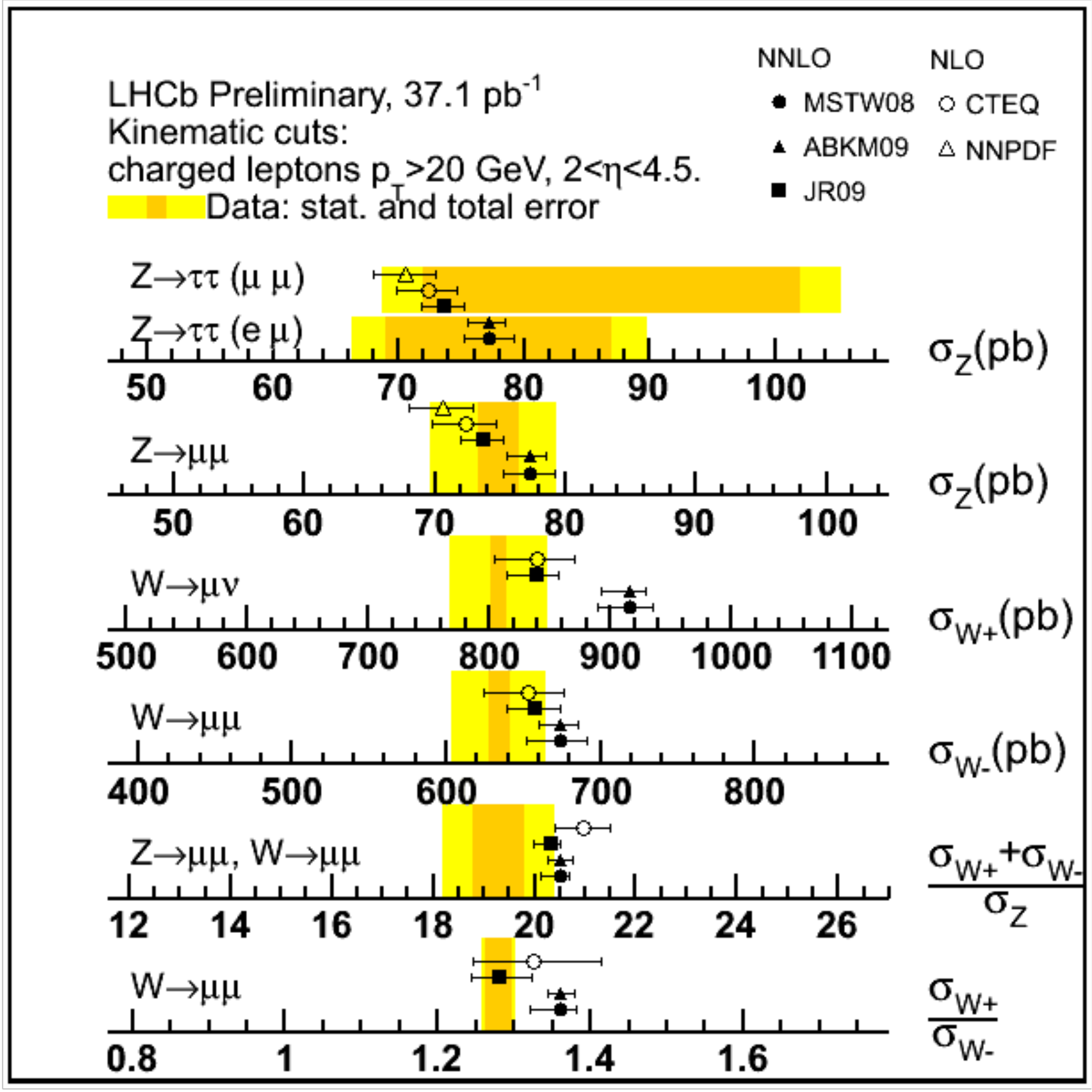}}
\caption{$Z$, $W^{+}$ and $W^{-}$ cross-section measurements and ratios shown as bands which illustrate the statistical (orange) and total (yellow) uncertainties. The measurements are compared to NNLO predictions with different proton PDF sets. \label{fig8}}
\end{figure}

The differential $Z\rightarrow\mu\mu$ cross-section in 5 bins of the boson rapidity and boson transverse momentum are shown in figure \ref{fig6} while the W charge asymmetry and $W^{+}$ to $W^{-}$ ratio, both in 5 bins of lepton pseudorapidity, are shown in figure \ref{fig7}. In the forward region, unlike for ATLAS and CMS, the W asymmetry distribution changes sign, showing the differing helicities of the couplings of the leptons. A summary of the measured cross-sections and the ratios $R_{W}$ and $R_{WZ}$ are shown in figure \ref{fig8}.

\section{Conclusions}
Measurements of the $W\rightarrow\mu\nu$ and $Z\rightarrow\mu\mu$ production cross-sections and their ratios in pp collisions at $\sqrt{s} = $ 7 TeV have been performed using $37.1 \pm 1.3$ pb$^{-1}$ of data collected by the LHCb detector. In addition, using 247 pb$^{-1}$ of data collected during 2010 and 2011, a measurement of the $Z\rightarrow\tau\tau$ cross-section has also been made. All measurements are in agreement with NNLO QCD predictions. Most of the errors are dominated by systematic uncertainties due to the determination from data of the various reconstruction efficiencies, with the larger statistics these systematic uncertainties will be reduced. Using the full 2011 dataset, it is expected that improved measurements will place significant constraints on the proton PDFs.

\end{document}